\documentclass[%
reprint,%
amssymb, amsmath,%
aip,pop,%
]{revtex4-2}

\usepackage{graphicx}
\usepackage{dcolumn}
\usepackage{bm}
\usepackage[mathlines]{lineno}
\usepackage{graphicx,times}
\usepackage{color}
\usepackage{amsfonts}
\usepackage[greek, british]{babel}
\usepackage{subfigure}
\usepackage{tabularx}
\usepackage{epsfig}

\begin{document}

	\title{Laser light scattering (LLS) to observe plasma impact on the adhesion of micrometer-sized particles to a surface}

	\author{D Shefer}
	\affiliation{Eindhoven University of Technology, Department of Applied Physics, Eindhoven, 5600 MB, The Netherlands}
	\email{d.shefer@tue.nl}
    \author{A Nikipelov}
    \affiliation{ASML, Veldhoven, 5504 DR, The Netherlands}
    \author{M van de Kerkhof}
    \affiliation{Eindhoven University of Technology, Department of Applied Physics, Eindhoven, 5600 MB, The Netherlands}
    \affiliation{ASML, Veldhoven, 5504 DR, The Netherlands}
    \author{V Banine}
    \affiliation{Eindhoven University of Technology, Department of Applied Physics, Eindhoven, 5600 MB, The Netherlands}
    \affiliation{ASML, Veldhoven, 5504 DR, The Netherlands}
    \author{J Beckers}
    \affiliation{Eindhoven University of Technology, Department of Applied Physics, Eindhoven, 5600 MB, The Netherlands}

\begin{abstract}

    Laser Light Scattering (LLS) method, combined with a long-distance microscope was utilized to detect micrometer-sized particles on a smooth substrate. LLS was capable to detect individual particle release, shrink, or fragmentation during exposure to a plasma or a gas jet. In-situ monitoring of hundreds of particles was carried out to investigate the effect of hydrogen plasma exposure on particle adhesion, morphology, and composition. LLS was calibrated with monodisperse melamine resin spheres with known sizes of 2.14~μm, 2.94~μm, and 5.26~μm in diameter. The lowest achievable noise level of approximately 3\% was demonstrated for counting 5.26~µm spherical melamine particles. The accuracy for melamine particle size measurements ranged from 50\% for 2.14~μm particles to 10\% for 5.26~μm particles. This scatter was taken as the imprecision of the method. Size distribution for polydisperse particles with known refractive index was obtained by interpolating to an effective scattering cross-section of a sphere using Mie theory. While the Abbe diffraction limit was about 2~μm in our system, the detection limit for Si particles in LLS according to Mie approximation was assessed to about 3~μm, given the limitations of the laser flux, microscope resolution, camera noise, and particle composition. Additionally, the gradual changes in forward scattering cross-sections for Si particles during the exposure to the hydrogen plasma were consistent with Si etching reported in the literature.

\end{abstract}
	

\maketitle

\section{Introduction}
    
    Under some conditions, plasma exposure is known to cause the release of nanometer and micrometer-sized particles from surfaces.\cite{Heijmans2016} Technologies sensitive to plasma-induced particle release are of special interest. For example, NASA’s study of the lunar and Mars surfaces confirmed suspended dust without settling.\cite{Horanyi1996, Sickafoose2002, AfsharMohajer2015} This effect is attributed to UV or plasma charging and may have a negative impact. For example, the mobility of micrometer-sized particles in plasma presents a challenge to solar panel longevity. In another example, a reticle (integrated circuit photo-mask), used in Extreme Ultraviolet (EUV) lithography is highly sensitive to contamination with particles of 20~nm and larger.\cite{vandeKerkhof2019, vandeKerkhof2018, Fu2019} Such particles may deposit on reticles even in the extremely clean environments of an EUV scanner in the presence of EUV-induced plasma.\cite{Beckers2019} Finally, in nuclear fusion plasma vessels (e.g. in ITER), plasma-facing walls releasing particles may deteriorate the gas mix. Because of tritium gas held in wall materials, dust generation in ITER is a serious concern, both from an erosion aspect and due to possible impurity release into the plasma.\cite{Tolias2016_fusion_plasma_contamination, Brezinsek2017} With respect to all these applications, the study of the behavior of micrometer-sized particles attached to a surface and interacting with plasma is important. To enable further studies, the development of new in-situ diagnostic tools is highly relevant.
    
    Traditionally used in the semiconductor industry, Laser Light Scattering (LLS) detects single particles on smooth or patterned substrates by analyzing light scattered into different angles from a relatively small illuminated spot (typically, around 10~µm).\cite{Germer2000, INOUE2002} Particles bigger than 1~µm scatter most of the light in the forward direction. Hence, a reflective substrate is a convenient method to improve such particle visibility. 

    With respect to the system of a particle attached to a surface, the particle adheres due to the combination of electrical, van der Waals (vdW), and capillary forces, as well as due to the particle’s chemical interaction with the surface. Adhesion depends on the particle’s size, composition, and morphology. A change in one of these parameters also affects the forward-scattered light intensity; hence, this can be used as a diagnostic method. In our work, we apply the LLS method, combined with long-distance microscopy, to image micrometer-sized particles. It will be demonstrated that the LLS method can be adapted in order to in-situ observe micrometer-sized particles on a surface placed in plasma or in other stressed conditions such as those caused by a gas jet. The advantage of the LLS method over traditional SEM measurement used in morphological diagnostics is the non-invasive in-situ manner of measuring which directly shows the impact of plasma treatment on particles during exposure.

\section{Apparatus and design}\label{Section-ii}
    

        Particles were deposited on the metallic side of the substrates; substrates used in all experiments were 1~inch in diameter polished sapphire wafers with 100~nm chromium coating. The mirror-finished wafers enable LLS to be operated in the dark field mode. The chromium coating is known to be robust against hydrogen embrittlement\cite{noauthor_1970-dw} and electrically conductive. The latter is necessary for SEM imaging before or after plasma exposure. Silicon (Si) particles were chosen in this work for the demonstration of the method because of the abundance of scientific literature on silicon including its etching by hydrogen plasma.\cite{Chang1982, Granata2013, Nazarov2008, Yoo2017} Melamine particles were selected because of their narrow standard deviation in size (when purchased commercially from Sigma Aldrich) and matte surface. Properties of the particles used in the experiments are listed in table~\ref{tab:calibration_samples}.
        
\begin{figure*}[t!]
        \centering
        \includegraphics[width=\linewidth]{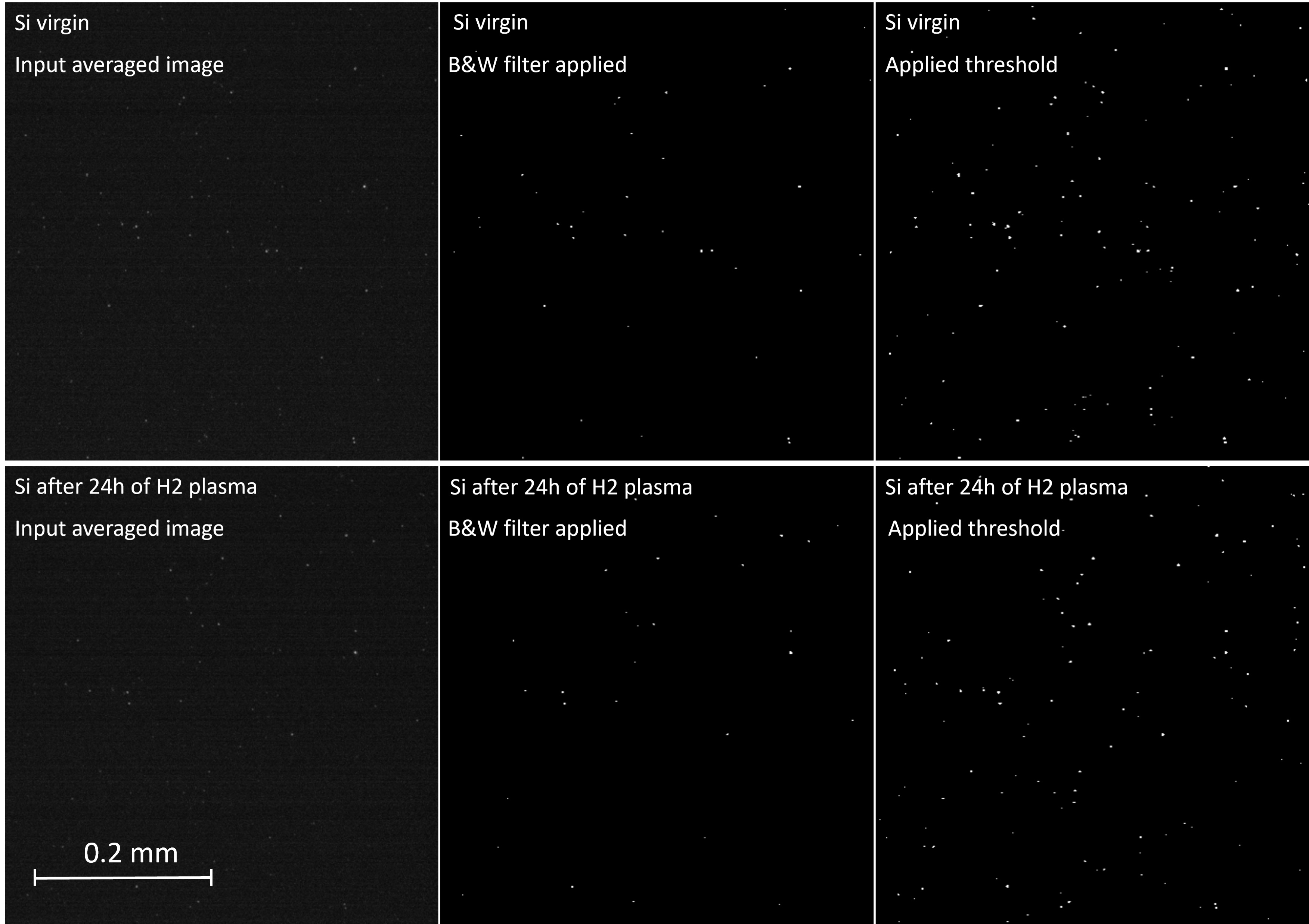}
        \caption{The acquired images of Si particles from the camera (virgin on the top and exposed to a hydrogen plasma for 6 hours on the bottom) and applied recognition filters.}
        \label{fig:grabbed_images}
    \end{figure*}

        The chromium substrates were contaminated with micrometer-sized particles using a Branson sonifier SFX 150 (40 kHz actuated tip). The sonifier disaggregated large clusters of particles by bringing its tip in contact with the edge of contaminated wafers. The average distance between the particles significantly exceeded their size (see Fig.~\ref{fig:grabbed_images}), which suppressed the effects of interference and simplified imaging, sizing of particles, and analysis of the interaction with plasma. 

        \begin{figure}[ht]
            \centering
            \includegraphics[width=\linewidth]{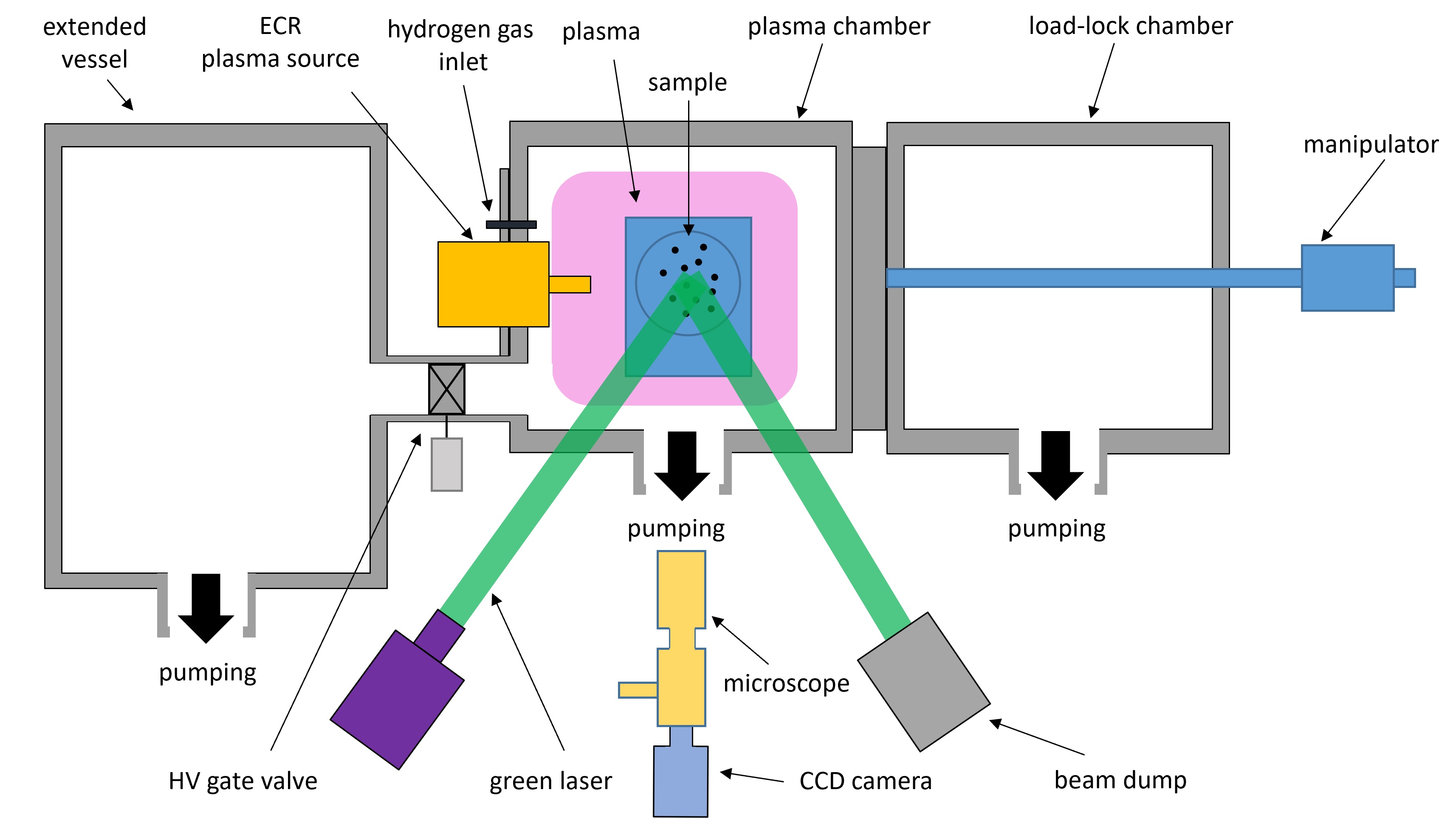}
            \caption{Schematic illustration of the used setup.}
            \label{fig:setup}
        \end{figure}

        A schematic overview of the used setup is depicted in Figure~\ref{fig:setup}. The setup comprised two vacuum chambers (a main chamber for the plasma and gas jet exposures and a load-lock chamber) separated by a VAT gate valve that remained closed during experiments. The main chamber was a 20x20x20~$cm^3$ cube with one of the flanges used for connection to the plasma source and the gas supply. A second flange of this chamber had an integrated window with an anti-reflective coating for LLS imaging. A third flange of this chamber was equipped with Philips vacuum gauges (HPT 200 Pirani/Bayard-Alpert and PPT 200 AR) which were both hydrogen calibrated. The flange with the plasma head also held a stainless steel wafer holder and allowed the swapping of wafers via the load-lock. The ultimate pressure in the vacuum chamber, achieved by a turbo-molecular pump (Pfeiffer THU 200 MP) and a scroll dry pre-pump (Edwards XDS10), was $10^{-4}$~Pa.

        During the experiments with plasma exposures, hydrogen was supplied to the main chamber at 30~sccm, resulting in a steady state pressure in the range of 1-10~Pa (mostly 5~Pa) without throttling the turbo-pump. The hydrogen plasma was driven by an Electron Cyclotron Resonance (ECR) plasma source (Aura-Wave, Sairem) at 100~W of RF power providing $T_e \simeq$~5~eV, $E_i \simeq$~15~eV, and ion flux toward the wafer of about $F \simeq$~1~A/$m^2$ according to Shirai~et~al\cite{Shirai_1989_ECR}. Under these conditions, the induced hydrogen radical ($H^{*}$) flux is expected to be 10 to 100 times higher than the $H^{+}$ flux due to a $\sim$10\% chance of $H^{*}$ association at the stainless steel walls of the main vacuum chamber compared to the 100\% chance of $H^{+}$ ion neutralization at the walls.\cite{Mozeti1999} Moreover, recombination of $H_{3}^{+}$ ions results in the generation of $\sim$2 radicals per event.\cite{Mendez2006} The selected conditions in this study featured a hundredfold more intense flux and approximately 5~times higher energy of ions compared to EUV-induced plasma\cite{vandeVen2018}. Hence, the exhibited results may be considered as the exposure to EUV plasma afterglow, accelerated at around 100~times.\cite{Kirchheim2014, vandeKerkhof2020_proc}
        
        \begin{figure}[ht]
            \centering
            \includegraphics[width=\linewidth]{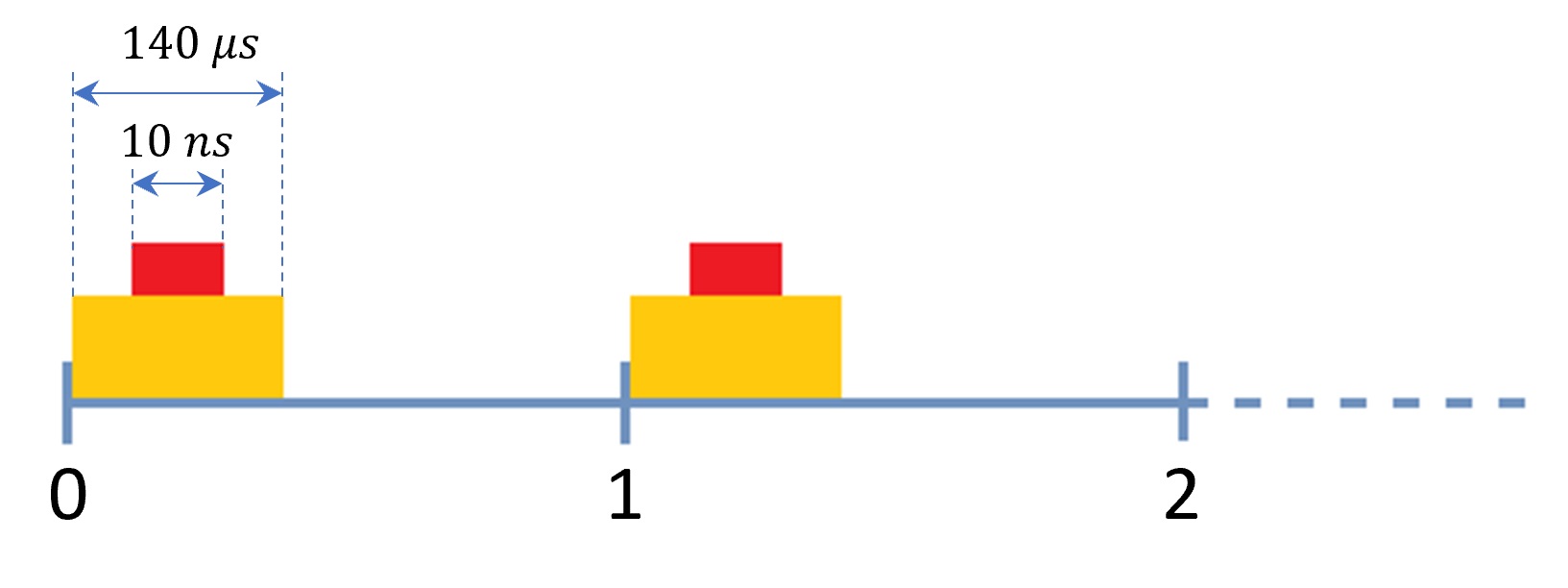}
            \caption{Schematic illustration of the synchronization timing between the camera and the laser system. The red parts represent the laser pulse durations and the orange parts represent the open time of the camera's shutter. Numbers 0, 1, 2 ..N indicate the laser pulses.}
            \label{fig:laser_camera_timing}
        \end{figure}

        For typical experiments a sample with particles was brought through the load-lock chamber to the middle of the main chamber (using a manipulator) and mounted vertically, facing the window with an anti-reflecting coating. A pulsed laser (EverGreen EVG00200, 70-200~mJ and 10~ns long pulses at 532~nm with 100-1000x attenuation by a grey filter), illuminated the wafer with a repetition rate of 0.71~Hz (1.4s between pulses). The laser beam, guided by mirrors, was expanded to 0.5~cm in diameter by two plano-convex lenses and entered the chamber through the window at about 10$^\circ$, reflected from the metal surface of the wafer, exited the chamber at 10$^\circ$, and was finally directed to a beam dump. The light scattered by particles on the surface was collected by a long-distance microscope (Distamax K2) with a working distance of 180~mm and a fully open aperture (diameter of 5~cm) with a CMOS camera (FLIR Grasshopper3) mounted to it. Pulsed laser illumination was chosen instead of illumination by a CW laser to reduce the blurriness caused by the vacuum pump-induced vibrations transferred to the microscope.

        The camera shutter was synchronized (Fig.~~\ref{fig:laser_camera_timing}) with the laser pulse by a signal delay generator (Model 577, BNC). Relatively short (140~µs) camera exposures helped to reduce the impact of the light from the plasma on the image background signal. The camera was configured to save 24-bit images with a resolution of 4,096~x~2,160 pixels. The pixel size was 3.45~x~3.45~µm$^2$, the quantum efficiency was 64\%, and the dynamic range was 65.15~dB. The maximal camera noise was 40.3~dB. The CMOS matrix size in combination with magnification by Distamax K2 and the distance to the sample (around 18~cm) produced a field of view (FoV) of 3~x~2~mm. This microscope FoV with a fully opened diaphragm was aligned with the illumination laser spot and the contaminated center of the wafer. The following camera settings were used: gain 48, gamma 0, black level 0, balance ratio 1.14, digital zoom - off, picture enhancer - off, full automatic control - off, auto exposure - off, auto white balance - off, black \& white compensation – off. The camera's gain had the greatest influence on the recognition of particles in post-processing steps.

        The acquired images were analyzed by a self-developed Python script. This script extracted the number of particles, their coordinates, and their total integrated intensities and sizes. The way for the size distribution of the particles was found using Mie theory is discussed below. To minimize the impact of laser beam power density fluctuations, the script applied a running average of 5 over the images, which was found to be an optimal value for the trade-off between the noise level and the time resolution achieved. The averaged total integrated scattering (TIS) of an image was computed by the script by summing the intensities of all pixels.

        The main chamber was also equipped with a flushing jet, which exhausted nitrogen gas pulses through a 4~mm tube placed at a 5~mm distance from the wafer and facing its center at 45$^\circ$. This flushing could be used to remove loosely bound particles from the substrate when the shear force exceeds the vdW force with which the particles are bound to the surface. The pulsed flushing was realized through a quick valve (DVI 005 M Pfeiffer) and a calibrated orifice (1.016~mm, Swagelok) that limited the flow. The pressure in the nitrogen line was measured by a Pfeiffer gauge (CPT 200 DN). The "flushing" jet could reach up to 6~nlm at the peak of the pulse. The main chamber had a bypass line to a volume extension vessel of 100 liters, separated from the main chamber by a VAT HV gate valve. During the flushing experiments, the turbo-pump was switched off and the bypass line was open. During plasma experiments, however, the bypass line remained closed. The extended vessel had its own pre-pump (Leybold SCROLLVAC 10). The sum productivity of the two pre-pumps for flushing experiments resulted in about 5~l/s at 100~Pa. The flushing pulses of 100~ms to 20~s were limited by the pre-pump productivity: long flushing pulses increased the pressure in the main chamber at the rate of 10~Pa in 10~s.

        In addition, to ensure the accuracy of the LLS setup calibration for measuring the sizes of silicon particles, a sample with silicon particles was additionally (measured on a similar, but not the same sample) qualified using SEM. The size distribution diagram obtained by SEM in a scanned area of 3x3 mm and analyzed by self-developed software was compared with the size distribution diagram obtained by LLS.

\section{Setup calibration}\label{Section-iii}
    
    The LLS technique enables monitoring of changes in the number of attached particles (Fig.~\ref{fig:grabbed_images}), as well as changes in the size distribution during exposure to plasma and flushing. The Figure shows the stages of image processing of Si particles before and after 6h of exposure to hydrogen plasma. The image clearly shows a change in the number of particles. In order to demonstrate the stability of the optical system, a seven-hour measurement of fixed-size particles (melamine) is used to calibrate the counting of particle numbers (see section~\ref{Section-iii-i}). Furthermore, a calibration for obtaining particle size distributions is performed based on Mie theory with a correction for the refractive index (see sections~\ref{Section-iii-ii} and~\ref{Section-iii-iii}). Finally, in section~\ref{Section-iii-iv} the calibration of the total substrate scattering will be demonstrated.

    \subsection{Particle number evaluation}\label{Section-iii-i}
    
        Evaluating the number of particles on the surface is challenging. For example, the resolution of the long-distance microscope is limited by the Abbe diffraction limit determined by the closest distance at which two separate sources of light can be distinguished from one another. This limit is expressed by\cite{Lipson1995-gi}
        
        \begin{eqnarray}\label{eq:1}
        d \approx \frac{\lambda}{2NA}
        \end{eqnarray}
        
        where $d$ is the minimum resolvable distance between two sources of scattered light, $\lambda$ is the wavelength of the laser light (532~nm) and NA is the numerical aperture (which in our configuration equals 0.137). Therefore, the resolution of our system is limited to approximately 1.9~μm.
        
\begin{figure}[ht]
            \centering
            \includegraphics[width=\linewidth]{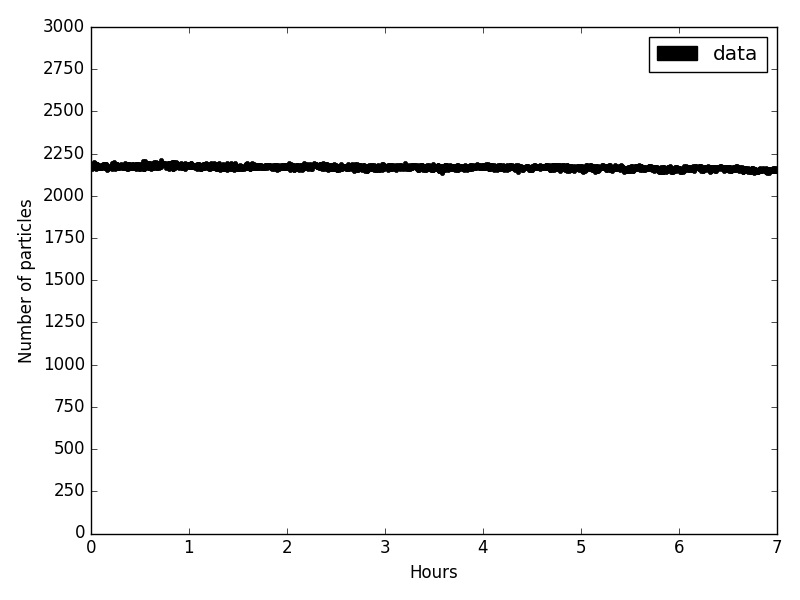}
            \caption{Calculated number of particles for a 7-hour-long camera baseline recorded for 5~μm melamine spheres (see table~\ref{tab:calibration_samples}).}
            \label{fig:baseline}
        \end{figure}

        The imaging of particles is limited not only by Abbe diffraction but also by the physical vibrations of the optical system, and variations of the particle shape and composition. In our experiments, the influence of camera noise, intensity fluctuations of the laser beam, and laser multimodality were also noted. Due to the limited coverage of these effects in the literature, comparisons were not made. Experimental uncertainties can be evaluated from measurements of scattering light from a stationary sample without disturbances. To enable this evaluation, a 7-hour-long imaging experiment of highly monodisperse 5.26~µm melamine spheres (see Table~\ref{tab:calibration_samples} with samples) was conducted. Note that in this experiment no flushing or plasma exposure was applied. The results (Fig.~\ref{fig:baseline}), demonstrate high laser stability and low counting uncertainty. In this experiment, the laser illumination and camera settings were identical to the experiments with plasma and flushing. It was shown that the dispersion of the number of detected particles was about 3\% (which is the lowest achievable noise level) with no long-term trends.

    \subsection{Size distribution of particles in LLS}\label{Section-iii-ii}

        Knowing the size distribution of processed particles is important. For instance, if large particles are more subjective to external stress factors, lowering their adhesion, such as those induced by exposure to plasma or a gas jet, the size distribution could shift toward smaller sizes. In another example, if exposure to plasma would lead to a developed surface and, thus, to a higher reflection coefficient of the incident light, the particles under the detection limit would become visible again. The particles that were already above the detection limit would shift toward larger sizes.

        \begin{table}[ht]
            \centering
            \begin{tabular}{ |p{0.3cm}|p{2.1cm}|p{1.25cm}|p{1.1cm}| }
            \hline
            № & Material & Size (μm) & SD (μm)\\
            \hline
            1 & melamine resin & 2.15 & 0.04\\
            2 & melamine resin & 2.94 & 0.05\\
            3 & melamine resin & 5.26 & 0.08\\
            4 & silicon & 5.00 & -\\
            \hline
            \end{tabular}
            \caption{Samples of particles used in the calibrations and experiments. Size is meant the diameter of the particle, and SD is short for standard deviation. Melamine particles were purchased from microParticles GmbH, silicon particles were purchased from US Research Nanomaterials, Inc.} \label{tab:calibration_samples}
        \end{table}
        
        The determination of the particle size distribution is even more complicated than the counting of particles. As generally known, CCD and CMOS cameras can be subjected to an effect called "blooming".\cite{Belloir2017} This blooming means that oversaturated pixels leak excess charge to their neighboring pixels. This process propagates until it reaches the edge, visibly and virtually enlarging the particle. Illumination of the entire particle requires sufficient illumination, and most of the particles under study scatter light in the flat Top-Hat regime, which means oversaturation of the pixels' capacity. Hence, the detected particle size as a number of bright pixels above the threshold is not consistent with the true particle size. A 2~μm particle occupied around 50 bright pixels (about 7 pixels in diameter) on the camera when in FoV. The only invariant in this problem is the integral of the photo-induced electrons in the camera's matrix or, in other words, the scattering efficiency of individual particles.

        Additional filtering must be applied before integrating the intensities of the pixels imaging the particles. After averaging the intensities of 5 images of 5 laser shots and applying the threshold value, the script filters tiny features (below 10 bright pixels in size). There are two reasons for this filtering. The first reason is that the high camera gain (max value, 48), used for high sensitivity, produces a few hot pixels that occur even without laser illumination and do not correspond to an actual signal. These hot pixels must be removed. The second reason relates to the presence of particles with sizes close to the detection limit. Due to the fluctuating laser intensity, these detections can appear and disappear from the detection region, significantly enhancing the noise level. Thus, by removing them, we focus on the residual population of particles that can always be identified with high confidence. 
        
\begin{figure}[ht]
            \centering
            \includegraphics[width=\linewidth]{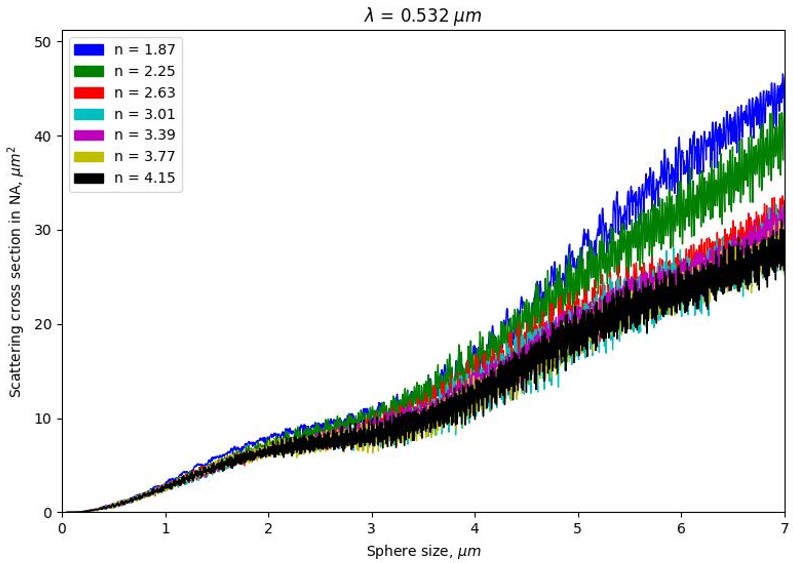}
            \caption{Plot of the results from the Mie scattering model as a function of different sizes and refractive indices (\emph{n}) scattered in the selected NA (corresponding to our optical system). The selected \emph{n} varied from 1.87 (melamine) to 4.15 (silicon).}
            \label{fig:mie_calculations}
        \end{figure}

        The correct approach would be to look at the scattering intensity of individual particles. As is generally known, particles of several micrometers in size obey Mie scattering theory.\cite{Horvath2009} The algorithm processing the collected images worked as follows. First, the scripted averaged intensities of 5 captured frames. Second, after applying the threshold, the intensities of images of the particles with an area larger than 10 pixels were integrated. Third, the scattering cross-section of the particle was calculated by multiplying the total intensity by the particle size with a constant, which is a fitting parameter to this model (see Eq.~2). Finally, an equivalent sphere with the same scattering cross-section and a refractive index was calculated using Mie theory, from which the size of the sphere/particle was derived. Therefore, measured scattering cross-sections can be translated into actual particle sizes using the Mie model for the light scattering by an individual particle. For this, a Mie calculator\cite{Wiscombe1979-dx} was used to evaluate the effective cross-sections of the particles for different particle sizes (from 0.1 to 7~µm). The absorption of light by the particles was not taken into account in the calculations due to a lack of available data. The results of the calculations for particles with a variety of refractive indices \emph{n} from 1.87 to 4.15 and the light collected in the NA corresponding to the microscope are plotted in Figure~\ref{fig:mie_calculations}. 
        
        In the Mie model, a spherical particle is situated in vacuum and emits light in all directions. Particles whose sizes are several times larger than the wavelength of the incident radiation predominantly scatter light forward and backward. We considered a model in which particles are positioned on a reflecting substrate, thus collecting only a portion of the forward and backward scattering into the NA of the microscope (NA = 0.137 for an objective lens with a diameter of 5 cm and a distance of 18 cm from the particles). It is worth noting that near-field effects due to reflection from the substrate were not taken into account. All calculations were performed assuming an isolated particle in vacuum with scattering confined to the chosen NA of the microscope.
        
        This graph shows that the particle's composition (i.e. the particles' refractive index) is more important for bigger sizes. Smaller particles are more sensitive to shape alterations. Our approach is to measure the scattering efficiency for the particles of known size and composition (in our case, monodisperse melamine spheres) as calibration. After this, for any material (i.e. refractive index) of interest, the cross-section of each particle can be translated into the size using the corresponding calibration curve from Figure~\ref{fig:mie_calculations}.
    
    \subsection{Effective scattering cross-section calibration}\label{Section-iii-iii}

\begin{figure}[ht]
            \centering
            \includegraphics[width=\linewidth]{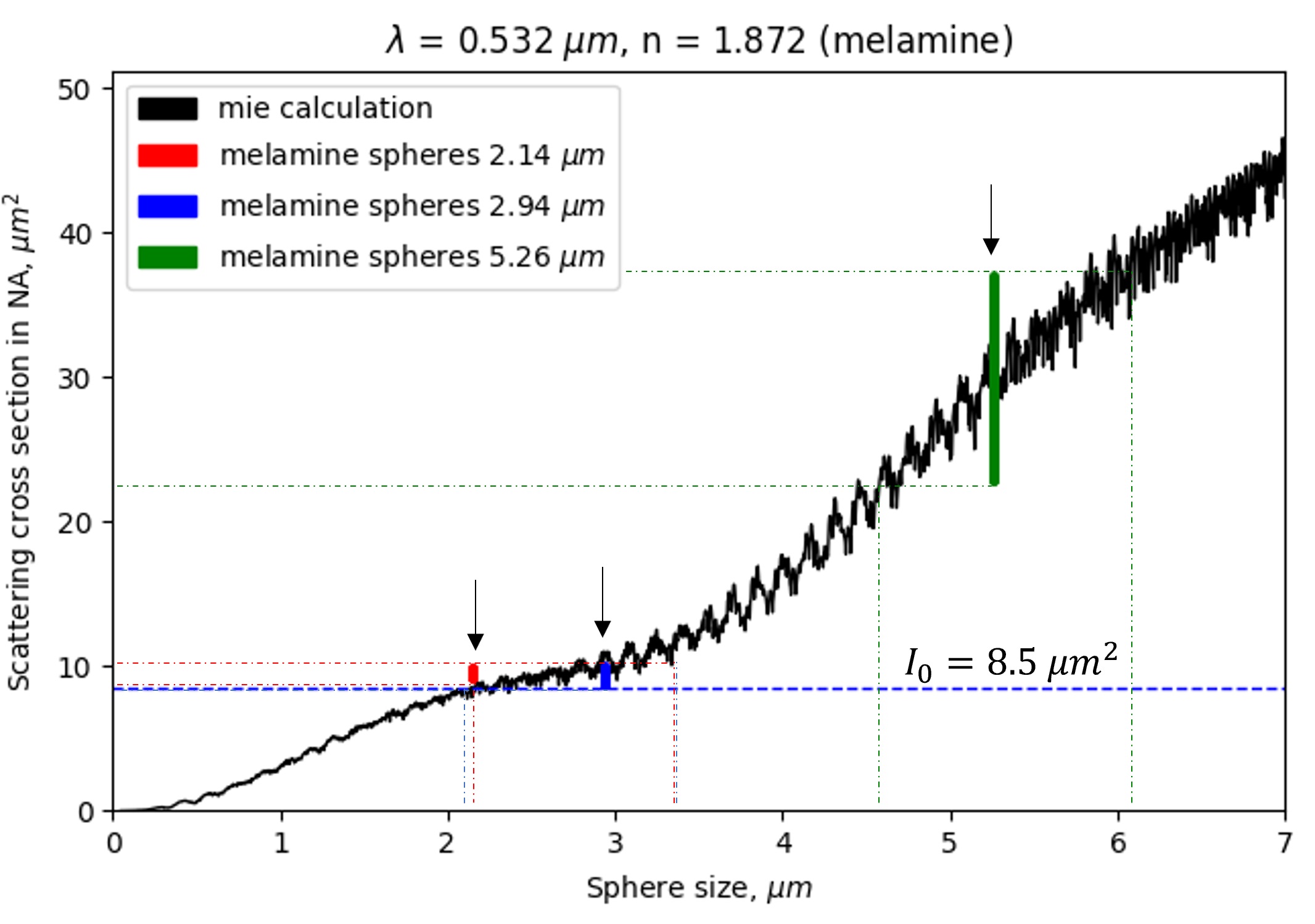}
            \caption{Calculated Mie scattering cross-section for a single melamine (n~=~1.872) particle depending on its size (black line). The arrows indicate the fitted measurements of the calibrated melamine particles from table~\ref{tab:calibration_samples} using formula~2. The horizontal blue dashed line indicates the $I_0$ value which is equal to 8.5~$\mu m^2$ and can be considered as the detection limit of this method.}
            \label{fig:scattering_efficiency}
        \end{figure}

        In order to use the curves from Figure~\ref{fig:mie_calculations}, they have to be calibrated. The measured intensities were fitted with the Mie curve. The results of this fit can be seen in Figure~\ref{fig:scattering_efficiency}. The arrows indicate the measured cross-sections. The blue dashed line indicates the $I_o$ value and can be considered as the detection limit of this method (it is attributed to the camera's noise which is of the same size as the min detected particles). The sizes of the particles were declared rather monodisperse, according to the manufacturer, with only a small standard deviation (see table~\ref{tab:calibration_samples}), while the measured intensities had some uncertainty. The scattering cross-sections of the melamine particles were fitted using the formula

        \begin{eqnarray}
        I_{ec} = (1 / \alpha)\cdot A \cdot I_m + I_o
        \end{eqnarray}\label{eq1} 

        where $I_{ec}$ is the effective scattering cross-section and $I_{m}$ is the particle intensity measured by LLS. The constant \emph{A} equals 700 and is related to the conversion of the laser intensity to the camera counts (or pixel counts). The constant $\alpha$ is the intensity correction factor. The applied laser intensity changed from 1x, to 14x and to 20x depending on the size of the particles,i.e. 2.14, 2.94, and 5.26~µm particles respectively. Therefore, for the purpose of laser intensity normalization, the intensity factor $\alpha$ was taken equal to 1, 14, and 20 for measurements on 5.26, 2.94, and 2.15~µm-particles respectively. The parameter $I_o$ remained constant for all fits and was taken equal to 8.5~$\mu m^2$. Physically it can be attributed to the losses of higher orders of diffraction, reflections from substrate asperities, and camera noise.

        The uncertainty of the cross-sections (and related to it the size uncertainty which was nominated by a supplier) can be considered as error bars of the method. For example, the determination of the size of the 2.14~µm particles has an uncertainty of about $\pm$1~µm which is 50\% of their size. It explains why 2.14 and 2.94~µm particles appear to have the same scattering cross-sections. At the same time, the determination of the size of the 5.26~µm particles has an uncertainty of about $\pm$0.5~µm which is only 10\% of their size.

    \subsection{Calibration of the total substrate scattering}\label{Section-iii-iv}

        In addition to the measurements of the number of particles and the particle size (distribution), another possibility is to look at the total integrated scattering from the field of view of the microscope. Technically, the summed and averaged intensity of all pixels is like an analog signal and, therefore, is more reliable as it avoids any image processing other than thresholding for noise removal.

        As mentioned, particles of several micrometers in size - as is the case here - obey Mie scattering theory: the scattered intensity is proportional to the particle cross-section (or to $r^2$ of the particle, where \emph{r} is the radius) and depends on multiple parameters such as \emph{n}, \emph{k} and \emph{D/$\lambda$}, and the polarization of the incident and collected light.\cite{Wiscombe1979-dx} For instance, melamine resins have n~=~1.872, k~=~0 (extinction coefficient is approximately zero for melamine-based materials in the visible range of wavelengths\cite{Kalyanaraman2014}), \emph{D/$\lambda$} is equal to 4.0, 5.5, 9.9 (for 2.14, 2.94 and 5.26~μm particles respectively). The incident light in our experiments was polarised perpendicular to the plane made up by the incoming beam, the reflecting beam, and the camera. The reflected light was not measured but expected to remain unchanged for particles significantly exceeding the wavelength of the radiation. A change in one of these parameters can be diagnosed by the TIS approach.
        
        The resolution limit of the TIS can be derived by matching it, again, with the Mie calculations for the given size, reflective index, and NA. The amount of scattering by a single particle was obtained by dividing the TIS by the number of detected particles of fixed size (melamine samples in table~\ref{tab:calibration_samples}). The sizes of the particles were taken according to the values declared by the manufacturer. The results of this calibration (Fig.~\ref{fig:tis_calibration}) show a perfect match with the previously calibrated scattering cross-sections which proves that imposed filtering, thresholding, and image processing used in the previous subsection do not contribute to the uncertainty in size determination significantly. The good match is explained by testing monodisperse spheres with low standard deviation. When applying the TIS signal for polydisperse particles, the match will be less good. Therefore, it can be concluded that the resolution of the TIS measurements and the effective scattering cross-section of individual particles is the same.

\section{Results for LLS measurements of silicon particles exposed to flushing and plasma}\label{Section-iv} 
    
    Silicon particles were exposed to a series of external stress factors such as flushing and plasma. The sequence of flushing-1 (10~min), plasma exposure (24~h), and flushing-2 (10~min) was applied to a wafer contaminated with Si particles. The flushing power was selected based on the median considerations. The flow must be strong enough to remove a noticeable amount of particles (exceeding the noise level of about 3\% as obtained in the calibration section). Physically, this would imply that the flushing shear force and the average adhesion force are comparable. Flushing removes particles, while adhesion keeps them in place. If a particle remains on the substrate after flushing, it means the adhesion force is equal to or greater than the removal force. The flushing (using nitrogen gas) used in the sequence consisted of 3-second long pulsed exhausts (6~nlm flow) at a frequency of 0.01~Hz (every 100~sec). Each flushing campaign lasted 10~min. Between two flushing campaigns, the samples were exposed to the hydrogen ECR plasma with the parameters described before. The quantification of the results used the calibrations described in the previous section.

    \begin{figure}[ht]
        \centering
        \includegraphics[width=\linewidth]{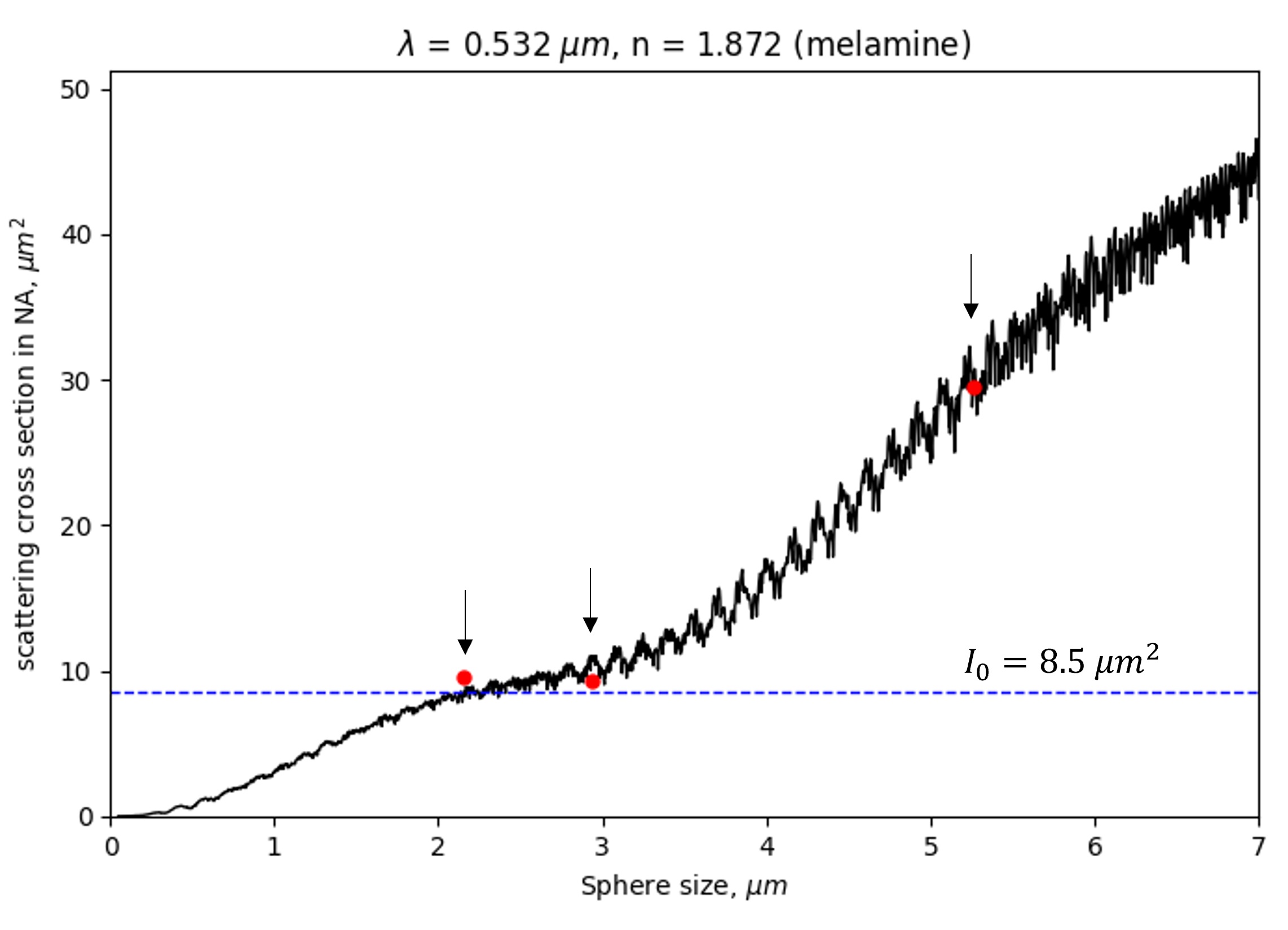}
        \caption{Calculated Mie scattering cross-section (black line) for a single melamine (\emph{n}~=~1.872) particle, depending on its size. The red dots (pointed by black arrows) indicate the averaged TIS measurements divided by the number of detected 2.14, 2.94, and 5.26~µm melamine particles from table~\ref{tab:calibration_samples}. The Mie scattering cross-section was fitted using the Eqn.~\ref{eq:1} with the same constants. The blue dashed line indicates the fitted constant $I_0$ which is equal to 8.5~µm$^2$ and can be considered as the detection limit of this method.}
        \label{fig:tis_calibration}
    \end{figure}
    
\begin{figure}[ht]
        \centering
        \includegraphics[width=\linewidth]{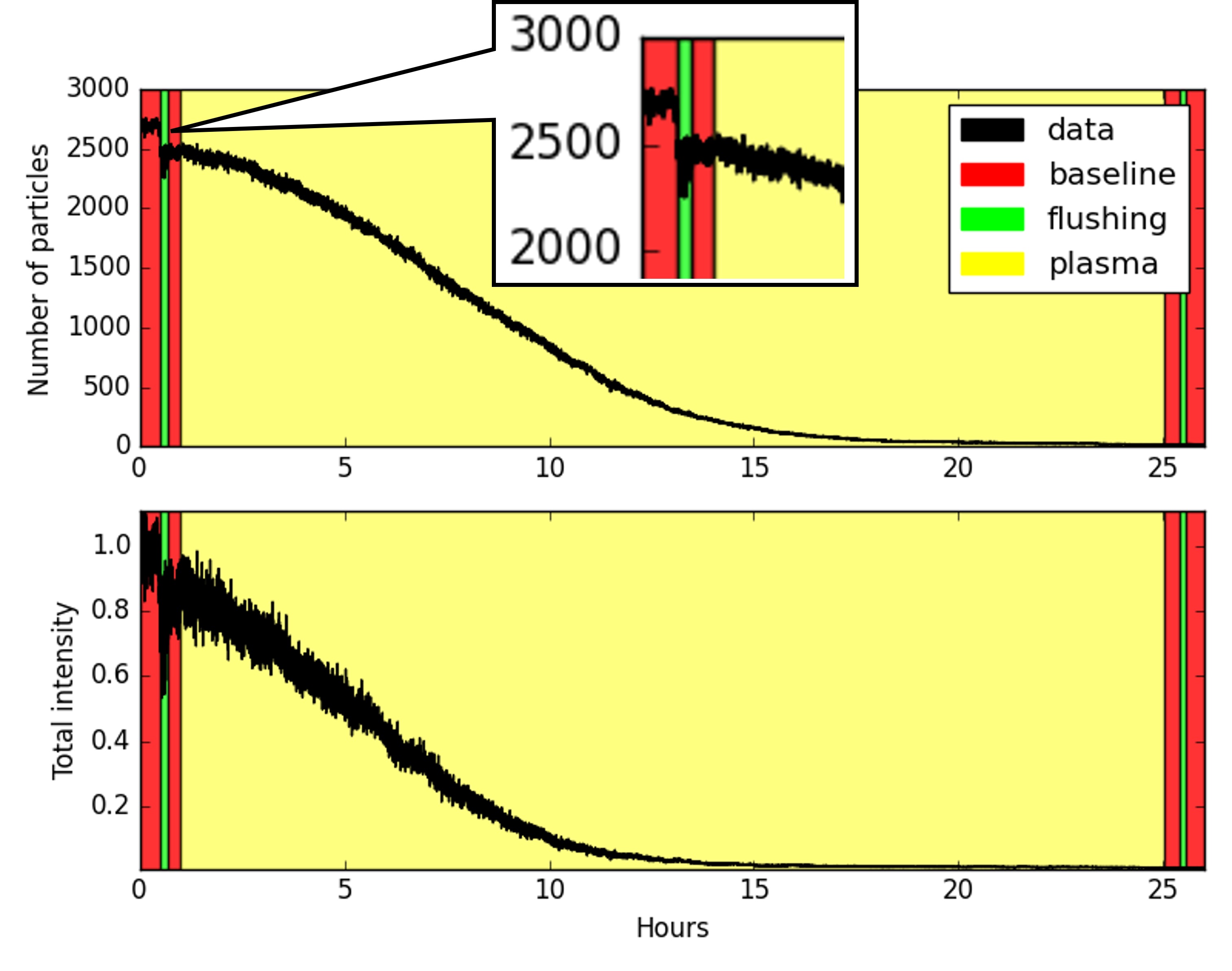}
        \caption{Calculated number of Si particle (on the top) and TIS signal (on the bottom) over time of exposure. Each dot is the average of 5 captured frames. particles above the detection limit.}
        \label{fig:si_24h_FPF_h2_100w}
    \end{figure}

    The top graph in Figure~\ref{fig:si_24h_FPF_h2_100w} shows the derived number of particles recorded over the experiment. The types of exposures (flushing or plasma) are mapped in different colors. Baselines (no exposures, only pressure changes) are shown in red, flushing campaigns are shown in green and the plasma exposure is shown in yellow. The plot shows that a significant amount of particles was flushed after the first few pulses. Further flushing appears to be ineffective, meaning that the remaining particles are attached with a force exceeding the applied shear force. The intermediate part of the experiment, during plasma exposure, clearly shows that the number of particles monotonically decays over the exposure which indicates the effect of plasma exposure on the particles' adhesion. This effect is the quantification of the impact shown in the grabbed images from the camera (Fig.~\ref{fig:grabbed_images}). The bottom graph in Figure~\ref{fig:si_24h_FPF_h2_100w} shows the TIS signal which correlates with the top graph and confirms that the intensity drop correlates with the number of scattering centers. The more rapid decay of the TIS signal compared to that of the number of particles during the first hour of plasma exposure needs more investigation. However, hypothetically, this effect could be explained by the presence of a native oxide shell or an adsorbed water layer around the particles that have different \emph{n} and \emph{k} (i.e. lower scattering), the oxide shell disappears after the first exposure to hydrogen plasma. After this phase, the scattering is proportional to the number of particles.

    The interpretation of the gradual decrease of Si particles during plasma exposure can be the following. First, upon plasma impact, a particle may develop asperities across its surface which reduces the effective vdW force which, in turn, promotes the specific particles to be released.\cite{Shefer2023} An alternative could be the weakening of the interfacing (binding) atomic layers mechanism, e.g. removal by plasma of intermediate adsorbate layers or removal of water forming hydrogen bridges.\cite{Kerkhof2020}. Another possible explanation could be the etching of the particles' material. The silane molecule $SiH_4$ is a formation product of sputtered Si atoms reacting with free hydrogen radicals, and it is volatile under our conditions. If the particles - due to this etching - shrink below the detection limit, they disappear from the sub-set of particles detected by the script, and the number of particles is reduced. The second flushing campaign was not necessary due to the lack of remaining measurable particles. Overall, these measurements show that the particles with the adhesion force exceeding the shear force during the first flushing campaign became loose due to plasma exposure. The results are consistent with literature data about the etching of silicon in hydrogen plasma.\cite{Chang1982, Granata2013, Nazarov2008, Yoo2017}

\begin{figure}[ht]
        \centering
        \includegraphics[width=\linewidth]{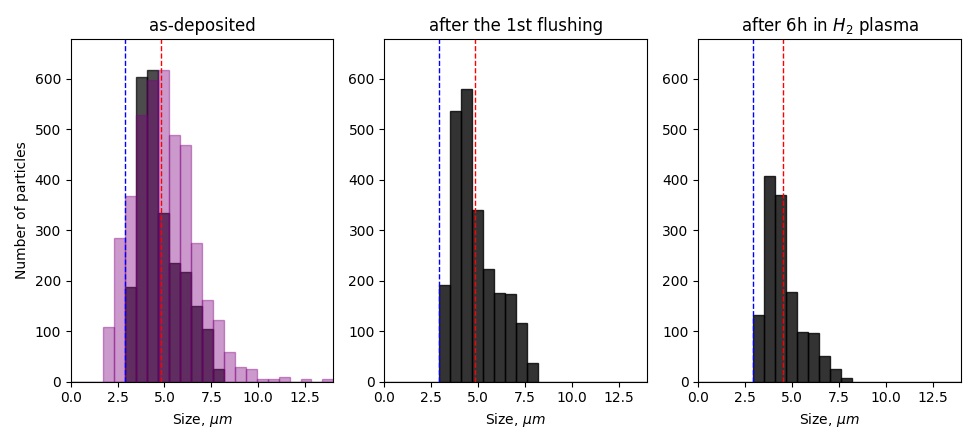}
        \caption{The black bins of the histograms indicate the number of Si particles measured by LLS after their deposition (on the left), after the first flushing campaign (in the middle), and after 6h of exposure to hydrogen plasma (on the right). The purple bins indicate the size distribution of Si particles measured in SEM after their deposition. The red dashed line indicates the mean value of the black bins. The blue dashed line indicates the detection limit of LLS in our system.}
        \label{fig:si_size_distribution}
    \end{figure}

    The histograms in Figure~\ref{fig:si_size_distribution} show the comparison of the size distributions of Si particles (black bins) after deposition (on the left), after the flushing (in the middle), and after 6h of $H_2$ plasma exposure (on the right). In addition, the size histogram obtained from SEM measurements on a similar (but not the same) sample with virgin Si particles (scanned over an area of 3~mm~x~3~mm ) is demonstrated in purple on the left "as-deposited" histogram for comparison. The particle size distribution histograms generated from in-situ laser light scattering (LLS) measurements were derived using the calibration procedure described above. The recorded intensities of Si particles were recalculated into sizes using the black curve from Figure~\ref{fig:mie_calculations} corresponding to silicon. The uncertainty of the method for these particles is the same as for melamine particles. The blue dashed line indicates the detection limit of the system which depends on \emph{n}. In fact, the detection limit is determined by the size, at which the constant $I_o$ intersects with the Mie calculation curve. For Si particles with \emph{n}~=~4.15, the detection limit is around 3~µm. 

    The histogram of as-deposited particles demonstrates the good matching of mean values in the calibrated LLS measurement results compared to size histograms obtained using SEM. The slight deviation in sizes is explained by the fact that SEM measurements were carried out for a similar sample with Si particles, but not the same (to prevent carbonization of particles in SEM and its influence on LLS measurements). It can also be seen from the plot, that after the first flushing a little fraction of the detected particles has been removed with no measurable difference in size distribution. Despite the fact that flushing scales as $d^2$ and adhesion should scale as $d$ we did not see the removal of bigger particles which can be addressed by the importance of other factors, such as size, shape, and roughness. As was mentioned before: this result indicates that the remaining particles have an adhesion force to the surface that exceeds the shear force exerted by the flushing. As is already shown in Figures~\ref{fig:grabbed_images} and~\ref{fig:si_24h_FPF_h2_100w}, the number of particles decays over the duration of hydrogen plasma exposure, while the histograms in Figure~\ref{fig:si_size_distribution} show that the particle size distribution has striven down and toward smaller sizes (together with the mean value shown as a red dotted line). As soon as a particle size reduces to the one indicated by the blue line (i.e. the detection limit), the particle disappears from the histogram, as it will not be detected anymore, and from the visibility of the script. 

    Therefore, the reliability of the recognition software has been tested based on 3 types of measurements: 
    \begin{enumerate}
        \item The stability of the number of particle detections was demonstrated in Figure~\ref{fig:baseline} for non-disturbed particles (without stressors like flushing or plasma) on a substrate.
        \item The reliability of the obtained size distribution is shown in Figure~\ref{fig:si_24h_FPF_h2_100w}, where the LLS measurements were compared to the SEM data (black bins vs purple bins).
        \item The average scattering cross-section of a melamine particle using the TIS signal was compared to individually detected particles and demonstrated a good match in Figures~\ref{fig:scattering_efficiency} and~\ref{fig:tis_calibration}. The TIS was treated as an analog signal for changing the scattering efficiency of particles.
    \end{enumerate}
    
    The obtained size histograms indicate that the etching mechanism with shrinking particles beyond the detection limit is the dominant mechanism for Si particle interaction with $H_2$ plasma. As can be seen from the middle and from right histograms, the highest percentage reduction was for the largest particles and the percentage gradually decreased toward the smallest particles. There are two reasons for that: 1) bigger particles shrink and take the place of smaller particles (hence, a relatively constant amount of small particles remained unchanged); 2) etching of Si by chemical sputtering of hydrogen radicals is only possible when accompanied by energetic electrons and ions from plasma breaking Si---Si bonds.\cite{Amuth2007} In that matter, the etching occurs at the place when particles interact with ions; hence, the particles are more to etch from the top rather than from the sides (it has been also demonstrated in AFM measurements\cite{Altmannshofer2016}). It explains why the entire histogram does not strive toward the smaller side as a whole. 
    
 \section{Conclusions}\label{Section-v}  

    The present study demonstrates the application of LLS, combined with long-distance microscopy, to in-situ characterize the response of micrometer-sized silicon particles on a smooth substrate to hydrogen plasma exposure or to a flushing gas jet. The number of particles, particle size distribution, and total scattering intensity (TIS) measured by laser light scattering (LLS) were calibrated with monodisperse melamine resin spheres. The results indicate that the counting accuracy was approximately 3\% for 5.26~µm melamine spheres. Furthermore, the observed inconsistency in relating the counting of only the bright pixels to the particle's size was attributed to the blooming effect. Therefore, Mie theory was applied to convert the calibrated particle effective scatter cross-sections to the size equivalent. The accuracy of the LLS size measurement was found to be between 50\% for 2.14~µm particles and 10\% for 5.26~µm particles.
    
    Surface-deposited Silicon particles were employed for LLS measurements in order to demonstrate the effectiveness of the method to serve as an in-situ diagnostic to visualize the effect of plasma exposure. The effect of plasma on Si particles is complex and may involve particle size and shape evolution due to chemical or physical sputtering. The in-situ measured counting and size evolution proves the etching of Si is dominant when exposed to $H_2$ plasma. The etching is mostly conducted by hydrogen ions. This is consistent with literature data obtained from SEM measurements. Additionally, SEM measurements conducted on virgin silicon particles demonstrated a high degree of concordance with the size distribution that was calculated using LLS and Mie theory and subsequently plotted.
    
    In conclusion, LLS can be useful as a tool for in-situ measurement of plasma exposure or gas jet flushing, fragmenting, or etching of micrometer-sized particles with a statistical description of adhesion for multiple (100-1000s) particles exposed to the same stressor.

\acknowledgments
    The assistance of P. Sanders, A. B. Schrader, J. T. Kohlhepp, and P. Minten in assembling the setup, as well as ASML in financial and scientific support, is gratefully acknowledged.

\bibliography{article_ref}

\end{document}